\RequirePackage{lineno}
\documentclass[aps,prc,showpacs,superscriptaddress,reprint]{revtex4-2}
\usepackage{rotating}
\usepackage{epsfig}
\usepackage[colorlinks=true, allcolors=blue]{hyperref}
\usepackage{lineno}

\usepackage{amsmath}
\usepackage{graphicx}
\usepackage[caption=false]{subfig}

\usepackage[dvipsnames]{xcolor}

\usepackage{xcolor}
\usepackage{xparse,xcoffins}

\ExplSyntaxOn
\NewCoffin\imagecoffin
\NewCoffin\labelcoffin

\keys_define:nn { fqwang/label }
 {
  label   .tl_set:N = \l_fqwang_label_tl,
  labelbox .bool_set:N = \l_fqwang_label_box_bool,
  labelbox .default:n = true,
  fontsize .tl_set:N = \l_fqwang_label_size_tl,
  fontsize .initial:n = \normalsize,
  pos .choice:,
  pos/nwHIGH .code:n = \tl_set:Nn \l_fqwang_label_pos_tl { left,upHIGH },
  pos/nwHigh .code:n = \tl_set:Nn \l_fqwang_label_pos_tl { left,upHigh },
  pos/nwhigh .code:n = \tl_set:Nn \l_fqwang_label_pos_tl { left,uphigh },
  pos/nw .code:n = \tl_set:Nn \l_fqwang_label_pos_tl { left,up },
  pos/nwlow .code:n = \tl_set:Nn \l_fqwang_label_pos_tl { left,uplow },
  pos/nwLow .code:n = \tl_set:Nn \l_fqwang_label_pos_tl { left,upLow },
  pos/nwLOW .code:n = \tl_set:Nn \l_fqwang_label_pos_tl { left,upLOW },
  pos/swHIHI .code:n = \tl_set:Nn \l_fqwang_label_pos_tl { left,downHIHI },
  pos/swHIGH .code:n = \tl_set:Nn \l_fqwang_label_pos_tl { left,downHIGH },
  pos/swHigh .code:n = \tl_set:Nn \l_fqwang_label_pos_tl { left,downHigh },
  pos/swhigh .code:n = \tl_set:Nn \l_fqwang_label_pos_tl { left,downhigh },
  pos/sw .code:n = \tl_set:Nn \l_fqwang_label_pos_tl { left,down },
  pos/swlow .code:n = \tl_set:Nn \l_fqwang_label_pos_tl { left,downlow },
  pos/swLow .code:n = \tl_set:Nn \l_fqwang_label_pos_tl { left,downLow },
  pos/swLOW .code:n = \tl_set:Nn \l_fqwang_label_pos_tl { left,downLOW },
  pos/n .code:n = \tl_set:Nn \l_fqwang_label_pos_tl { hc,up },
  pos/w .code:n = \tl_set:Nn \l_fqwang_label_pos_tl { left,vc },
  pos/s .code:n = \tl_set:Nn \l_fqwang_label_pos_tl { hc,down },
  pos/e .code:n = \tl_set:Nn \l_fqwang_label_pos_tl { right,vc },
  pos .initial:n = nw,
  unknown .code:n   = \clist_put_right:Nx \l_fqwang_label_clist
                       { \l_keys_key_tl = \exp_not:n { #1 } }
 }

\clist_new:N \l_fqwang_label_clist
\box_new:N \l_fqwang_label_box
\box_new:N \l_fqwang_label_image_box

\NewDocumentCommand{\xincludegraphics}{O{}m}
 {
  \group_begin:
  \tl_clear:N \l_fqwang_label_tl
  \clist_clear:N \l_fqwang_label_clist
  \keys_set:nn { fqwang/label } { #1 }
  \tl_if_empty:NTF \l_fqwang_label_tl
   {
    \fqwang_includegraphics:Vn \l_fqwang_label_clist { #2 }
   }
   {
    \SetHorizontalCoffin\imagecoffin
     {
      \fqwang_includegraphics:Vn \l_fqwang_label_clist { #2 }
     }
    \SetHorizontalCoffin\labelcoffin
     {
      \raisebox{\depth}
       {
        \bool_if:NTF \l_fqwang_label_box_bool
         { \fcolorbox{white}{white}{\l_fqwang_label_size_tl\l_fqwang_label_tl} }
         { \l_fqwang_label_size_tl\l_fqwang_label_tl }
       }
     }
    \SetVerticalPole\imagecoffin{left}{36pt+\CoffinWidth\labelcoffin/2}
    \SetVerticalPole\imagecoffin{right}{\Width-36pt-\CoffinWidth\labelcoffin/2}
    \SetHorizontalPole\imagecoffin{upHIGH}{\Height+6pt-\CoffinHeight\labelcoffin/2}
    \SetHorizontalPole\imagecoffin{upHigh}{\Height-0pt-\CoffinHeight\labelcoffin/2}
    \SetHorizontalPole\imagecoffin{uphigh}{\Height-6pt-\CoffinHeight\labelcoffin/2}
    \SetHorizontalPole\imagecoffin{up}{\Height-12pt-\CoffinHeight\labelcoffin/2}
    \SetHorizontalPole\imagecoffin{uplow}{\Height-18pt-\CoffinHeight\labelcoffin/2}
    \SetHorizontalPole\imagecoffin{upLow}{\Height-24pt-\CoffinHeight\labelcoffin/2}
    \SetHorizontalPole\imagecoffin{upLOW}{\Height-30pt-\CoffinHeight\labelcoffin/2}
    \SetHorizontalPole\imagecoffin{downHIHI}{36pt+\CoffinHeight\labelcoffin/2}
    \SetHorizontalPole\imagecoffin{downHIGH}{30pt+\CoffinHeight\labelcoffin/2}
    \SetHorizontalPole\imagecoffin{downHigh}{24pt+\CoffinHeight\labelcoffin/2}
    \SetHorizontalPole\imagecoffin{downhigh}{18pt+\CoffinHeight\labelcoffin/2}
    \SetHorizontalPole\imagecoffin{down}{12pt+\CoffinHeight\labelcoffin/2}
    \SetHorizontalPole\imagecoffin{downlow}{6pt+\CoffinHeight\labelcoffin/2}
    \SetHorizontalPole\imagecoffin{downLow}{0pt+\CoffinHeight\labelcoffin/2}
    \SetHorizontalPole\imagecoffin{downLOW}{-6pt+\CoffinHeight\labelcoffin/2}
    \use:x{\JoinCoffins\imagecoffin[\l_fqwang_label_pos_tl]\labelcoffin[vc,hc]} 
    \TypesetCoffin\imagecoffin
   }
   \group_end:
 }
\NewDocumentCommand{\setlabel}{m}
 {
  \keys_set:nn { fqwang/label } { #1 }
 }
\cs_new_protected:Nn \fqwang_includegraphics:nn
 {
  \includegraphics[#1]{#2}
 }
\cs_generate_variant:Nn \fqwang_includegraphics:nn { V }

\ExplSyntaxOff

\newcommand{\gevc}  {GeV/$c$}
\newcommand{\gevcc} {GeV/$c^2$}
\newcommand{\ns}    {n_5/s}
\newcommand{\minv}  {m_{\rm inv}}
\newcommand{\vpi}   {v_{2}}
\newcommand{\pair}  {{\rm pair}}
\newcommand{\vpair} {v_{2,\pair}}
\newcommand{\res}   {{\rm res}}
\newcommand{\vres}  {v_{\rm 2,\res}}
\newcommand{\vrho}  {v_{2,\rho}}
\newcommand{\os}    {{\rm os}}
\newcommand{\sm}    {{\rm ss}}
\newcommand{\vos}   {v_{2,\os}}
\newcommand{\vsm}   {v_{2,\sm}}

\newcommand\mean[1]{\left\langle#1\right\rangle}

\begin{document}

\title{Influence of the chiral magnetic effect on particle-pair elliptic anisotropy}

\author{Han-Sheng Li}
\email{li3924@purdue.edu}
\affiliation{Department of Physics and Astronomy, Purdue University, West Lafayette, IN 47907, USA}
\author{Yicheng Feng}
\email{feng216@purdue.edu}
\affiliation{Department of Physics and Astronomy, Purdue University, West Lafayette, IN 47907, USA}
\author{Fuqiang Wang}
\email{fqwang@purdue.edu}
\affiliation{Department of Physics and Astronomy, Purdue University, West Lafayette, IN 47907, USA}
\affiliation{School of Science, Huzhou University, Huzhou, Zhejiang 313000, China}

\begin{abstract}

Chiral Magnetic Effect (CME) is a phenomenon in which electric charge is separated by a strong magnetic field from local domains of chirality imbalance 
in quantum chromodynamics. The CME-sensitive, azimuthal correlator difference $\Delta\gamma$ between opposite-sign (OS) and same-sign (SS) charged hadron pairs is contaminated by a major physics background proportional to the particle elliptic anisotropy ($v_2$). The CME signal, on the other hand, contributes to the difference in the pair elliptic anisotropies between OS and SS pairs ($\Delta\vpair$). We investigate $\Delta\vpair$ and found its sensitivity to CME to be similar to that of the $\Delta\gamma$ observable.
\end{abstract}
\maketitle

\section{Inroduction}
One of the unsettled questions in relativistic heavy ion collisions is the chiral magnetic effect (CME). It is predicted by quantum chromodynamics (QCD) to exist because of vacuum fluctuations of the topological gluon field, yielding a chirality imbalance of (anti-)quarks in local domains because of quark-gluon interactions. Such a chirality imbalance would produce an electric charge separation under a strong magnetic field because of the charge-dependent magnetic moment of the (anti-)quarks~\cite{Kharzeev:1998kz,Kharzeev:2004ey,Kharzeev:2007jp,Fukushima:2008xe}. A strong magnetic field is presumably produced in non-head-on relativistic heavy ion collisions~\cite{Skokov:2009qp,Deng:2012pc}. However, definite signals of the CME have not yet been observed~\cite{Kharzeev:2015zncReviewCME,Zhao:2019hta}.

The magnetic field created in heavy ion collisions is on average perpendicular to the reaction plane (RP), the plane spanned by the beam and the impact parameter direction of the collision. The CME charge separation signal is an excess of positively charged particles along one direction perpendicular to the RP and an excess of negatively charged particles in the opposite direction~\cite{Kharzeev:2007jp}. It is convenient to express it in Fourier series~\cite{Voloshin:2004vk},
\begin{equation}
    dN_\pm/d\phi \propto 1 + 2v_1\cos\phi + 2v_2\cos2\phi + 2a_1^{\pm}\sin\phi + \cdots\,,
\end{equation}
where $\phi$ is the azimuthal angle of the particle momentum vector with respect to RP and the subscript $\pm$ indicates particle charge sign. The $v_n \equiv \mean{\cos n\phi}$ parameters are called flow harmonics and are taken as charge independent.
The charge-dependent $a_{1\pm}$ parameters characterize the CME signals in positive and negative charged particles, respectively; they are random in sign because of random fluctuations of the vacuum topological charge sign, rendering a vanishing average signal on single particle level~\cite{Kharzeev:2004ey}. The commonly used observable is thus a two-particle correlator~\cite{Voloshin:2004vk},
\begin{equation}
    \gamma=\mean{\cos(\phi_\alpha+\phi_\beta)}\,,
\end{equation}
where $\phi_\alpha$ and $\phi_\beta$ are the azimuthal angles of two particles of interest (POI).
Because of the presence of charge-independent background, such as effects from global momentum conservation, the difference between opposite-sign (OS) and same-sign (SS) correlators is used in experimental search for the CME~\cite{Voloshin:2004vk},
\begin{equation}
    \Delta\gamma \equiv \gamma_\os - \gamma_\sm\,.
\end{equation}
The CME signal presented in the $\Delta\gamma$ observable is $2a_1^2$~\cite{Jiang:2016wve}.
However, charge-dependent background is also present in the $\Delta\gamma$ correlator, such as correlations between daughter particles from a resonance decay, or among particles from the same jet or a back-to-back dijet~\cite{Voloshin:2004vk,Wang:2009kd,Liao:2010nv,Schlichting:2010qia}. This background correlation can be schematically expressed as~\cite{Voloshin:2004vk}
\begin{equation}
    \Delta\gamma_{\rm res} = \mean{\cos(\phi_\alpha+\phi_\beta)}\vres\,,
\end{equation}
where the subscript ``res'' stands generically for genuine two-particle correlation background such as those due to resonance decays, jets, quantum statistics, and $\vres\equiv\mean{\cos2\phi_\res}$ is the elliptic flow anisotropy of those two-particle correlation clusters.

The charge-dependent background turns out to be significant~\cite{Schlichting:2010qia}, and can nearly explain the measured $\Delta\gamma$ magnitude~\cite{Abelev:2009ac,Abelev:2009ad}. Large efforts have since been investigated to eliminate or mitigate this background~\cite{Adamczyk:2013kcb,Adamczyk:2014mzf,Khachatryan:2016got,Acharya:2017fau,STAR:2019xzd}, including innovative observables~\cite{Xu:2017qfs,Voloshin:2018qsm,STAR:2021pwb,Tang:2019pbl}.

Although the CME is a parity-odd phenomenon, the two-particle $\Delta\gamma$ correlator observable is intrinsically parity even. As a result, the CME signal will appear as a component in the pair-wise elliptic anisotropies ($\vpair$). It has been recently pointed out by Ref.~\cite{Xu:2023elq} that resonance $\vres$ and particle pair $\vpair$ contain CME information and should in principle be able to probe the CME as well. In this article, we follow up on this point and investigate further insights, if any, that pair $\vpair$ may bring to the CME search endeavor. 

\section{AVFD Simulation}
We use the Anomalous-Viscous Fluid Dynamics (AVFD) model~\cite{Shi:2017cpu,Jiang:2016wve,Shi:2019wzi} for our investigation. AVFD is an anomalous fluid dynamics developed to describe the evolution of chiral fermion currents in the quark-gluon plasma (QGP) created in relativistic heavy ion collisions in addition to the iEBE-VISHNU simulations. The iEBE-VISHNU package is a hybrid approach to event-by-event simulations of relativistic heavy-ion collisions based on (2+1)-dimensional viscous hydrodynamics coupled to a hadronic cascade model~\cite{Shen:2014vra,Heinz:2015arc}. 
The average magnetic field in AVFD is calculated from the spectator protons and it is along the  direction perpendicular to RP. 
In the AVFD model, the magnetic field is quantified by event-by-event simulations~\cite{Bloczynski:2012en} taking into account the fluctuations of the magnetic field direction with respect to the RP.
A modest time evolution is assumed for the decreasing magnetic field, $B\propto[1+(\tau/\tau_B)^2]^{-1}$, with a typical lifetime $\tau_B=0.6$~fm/$c$ comparable to the initial time of hydrodynamic evolution~\cite{Jiang:2016wve}.
The initial condition for the axial charge density ($n_5$) is dynamically generated in AVFD to be proportional to the entropy density ($s$), and the strength is set via the proportionality coefficient $\ns$.
The QGP medium evolution is simulated by viscous hydrodynamics to describe the bulk background in heavy ion collisions, with transport parameters for the diffusion coefficient as well as the relaxation time. 
A final hadronic stage after the hydrodynamic freeze-out is included with hadronic re-scatterings and resonance decays.

Au+Au collisions at 200 GeV were simulated by AVFD. In this study we focus on the same centrality range of 30-40\% as studied in Ref.~\cite{Choudhury:2021jwd}. 
Three $\ns$ values are used in the simulations, $n_{5}/s$=0, 0.1, and 0.2. Each of these simulations has around 20M events generated. 

In this work, the RP is taken from the simulation, fixed to zero. The POI's are taken to be within the acceptance of $|\eta|<1$ and $0.2 < p_{T} < 2$~\gevc, typical of midrapidity detectors such as the STAR experiment~\cite{Abelev:2008ab}.

\section{Resonance Elliptic Anisotropies}
\begin{figure}[!hbt]
    \xincludegraphics[width=0.8\linewidth,pos=swHIGH,label=\hspace{0.8cm}a)]{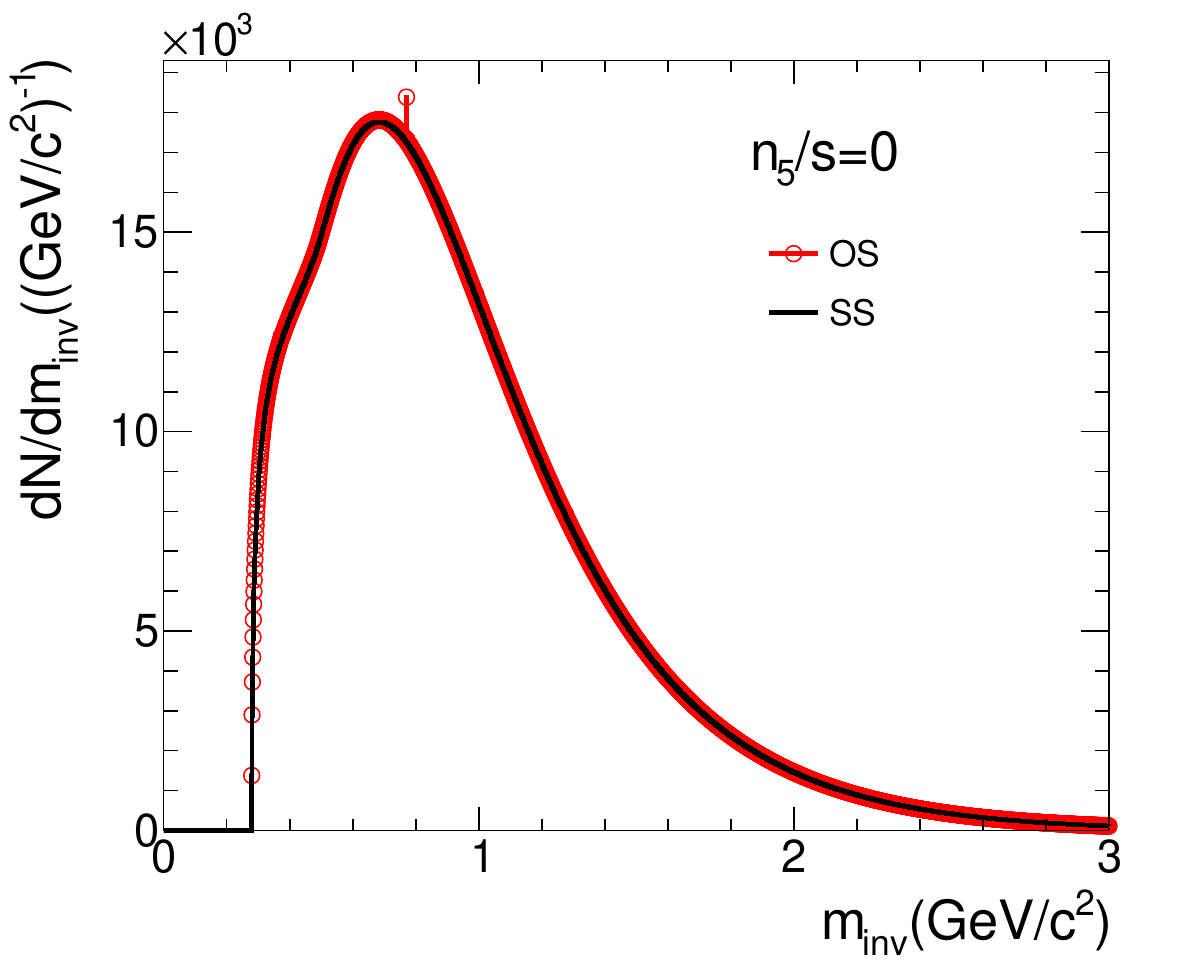}\hfill
    \xincludegraphics[width=0.8\linewidth,pos=swHIGH,label=\hspace{0.8cm}b)]{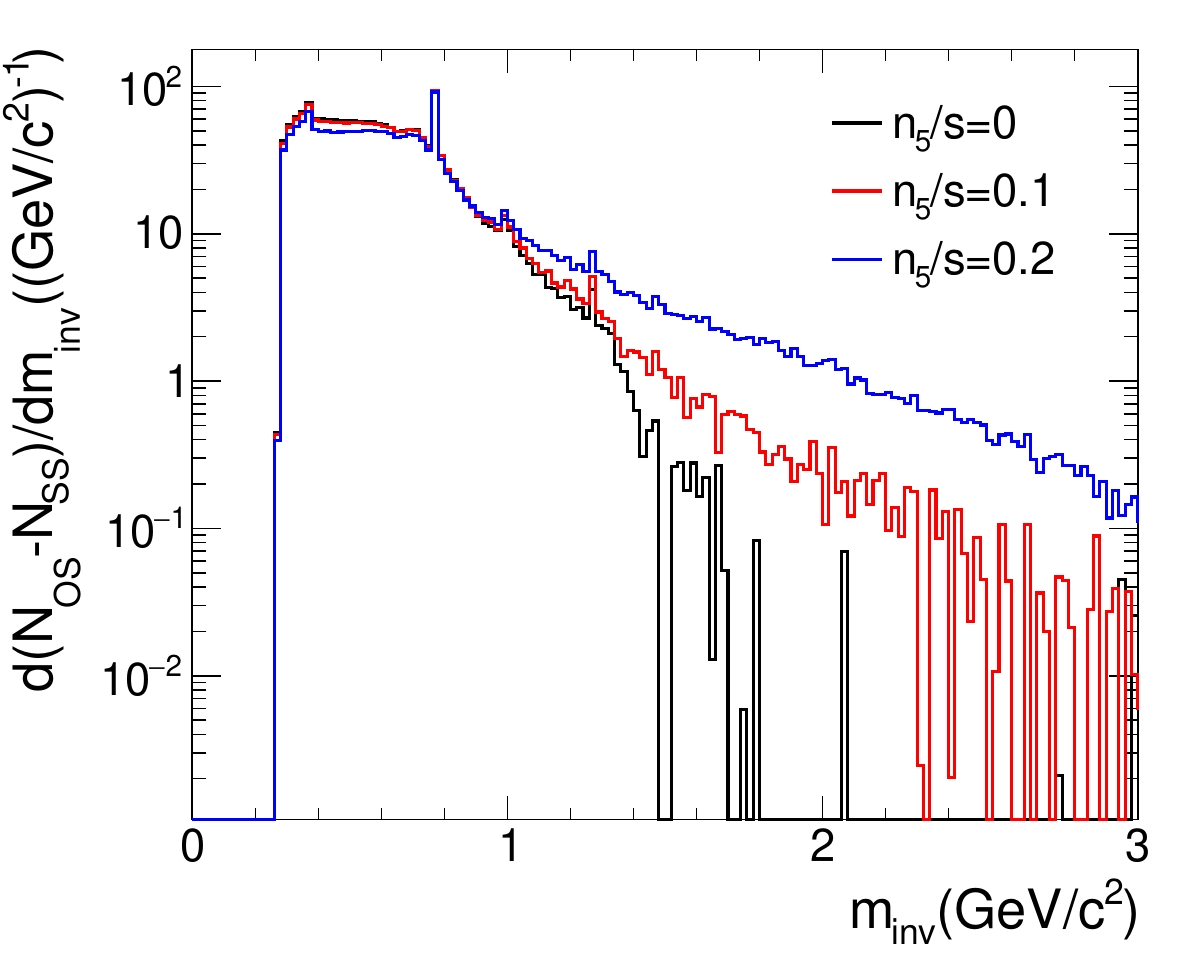}\hfill
    \caption{{(a) Invariant mass ($\minv$) distributions of opposite-sign (OS) pairs and same-sign (SS) pairs simulated by AVFD with axial current density of $\ns=0$.} Pion mass is assumed for all charged particles in calculating $\minv$. 
    (b) Difference in the $\minv$ distributions of OS and SS pairs from AVFD simulations with $\ns=0$, 0.1, and 0.2.} 
    \label{fig:minv}
\end{figure}

We examine the invariant mass ($\minv$) distributions of charged hadron pairs in this study. 
Figure~\ref{fig:minv}(a) shows the OS and SS pair $\minv$ distributions, assuming pion mass for all hadrons, from AVFD simulation with $\ns=0$ as an example. Figure~\ref{fig:minv}(b) shows the difference of the $\minv$ distributions between OS and SS pairs for all three $n_{5}/s$ values. The CME implemented in AVFD affects the relative OS and SS pair $\minv$ distributions--the distribution flattens with larger $\ns$. This arises from contributions from back-to-back OS pairs (hence larger $\minv$) and near-side SS pairs (hence smaller $\minv$), resulting in an enhancement at large $\minv$ and depletion at small $\minv$ for large $\ns$ compared to small one.

Several resonance peaks are apparent in the $\minv$ distributions, $\rho$, $f_0(980)$, and $f_2(1270)$. No intrinsic mass width is implemented in AVFD, so the resonances appear as sharp peaks. 
We note that there is a step at $\minv\sim 0.4$~\gevcc; such a step is not present in $\pi^+\pi^-$ invariant mass spectrum and is apparently caused by other pairs with the wrong pion mass used in the $\minv$ calculation. We use all charged hadrons in our AVFD study similar to experimental data analysis, but it would be desirable to use identified pions in the future.

\begin{figure*}[!hbt]
\xincludegraphics[width=0.333\linewidth,pos=nwlow,label=\hspace{3.5cm}a)]{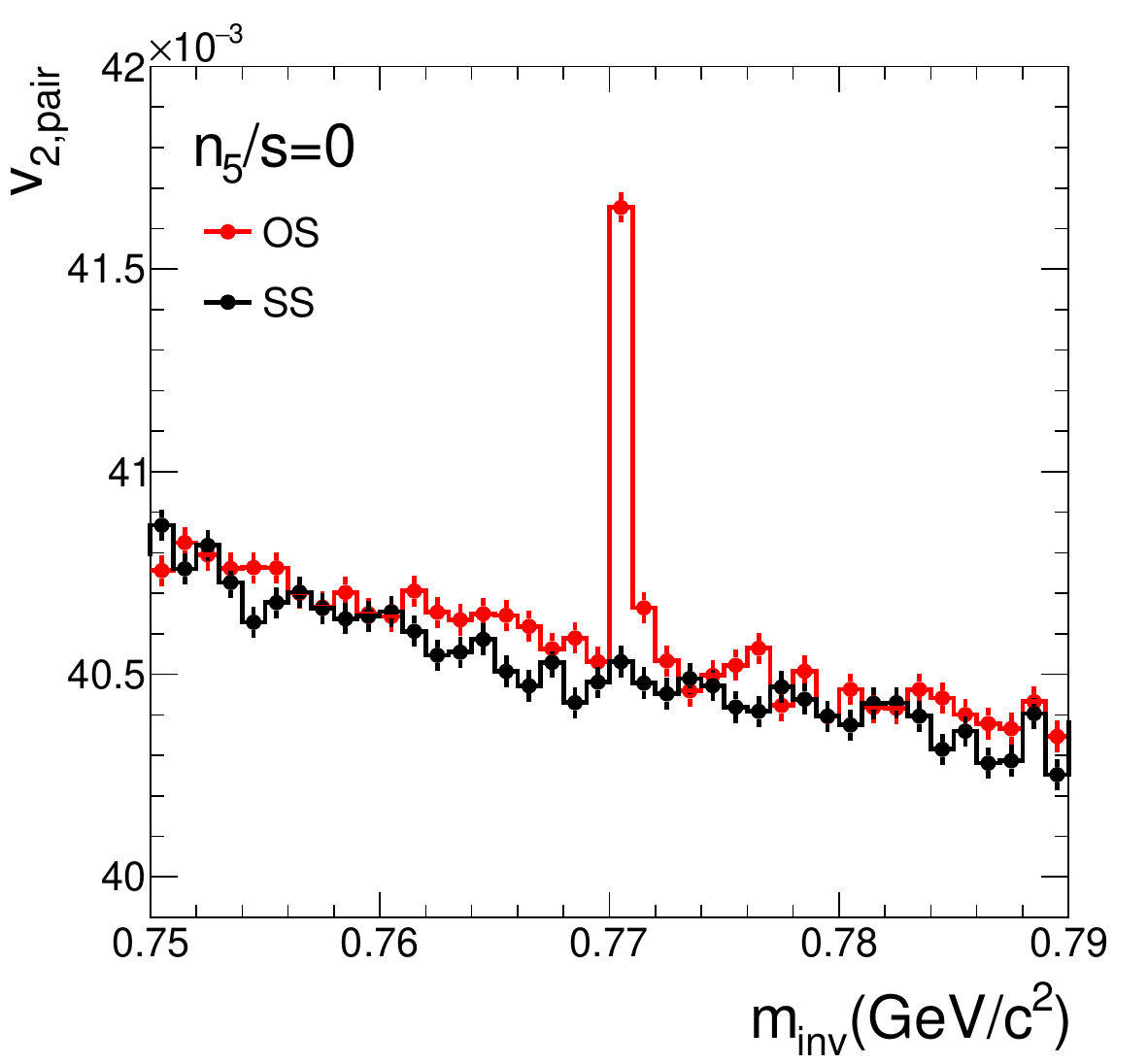}\hfill
\xincludegraphics[width=0.333\linewidth,pos=nwlow,label=\hspace{3.5cm}b)]{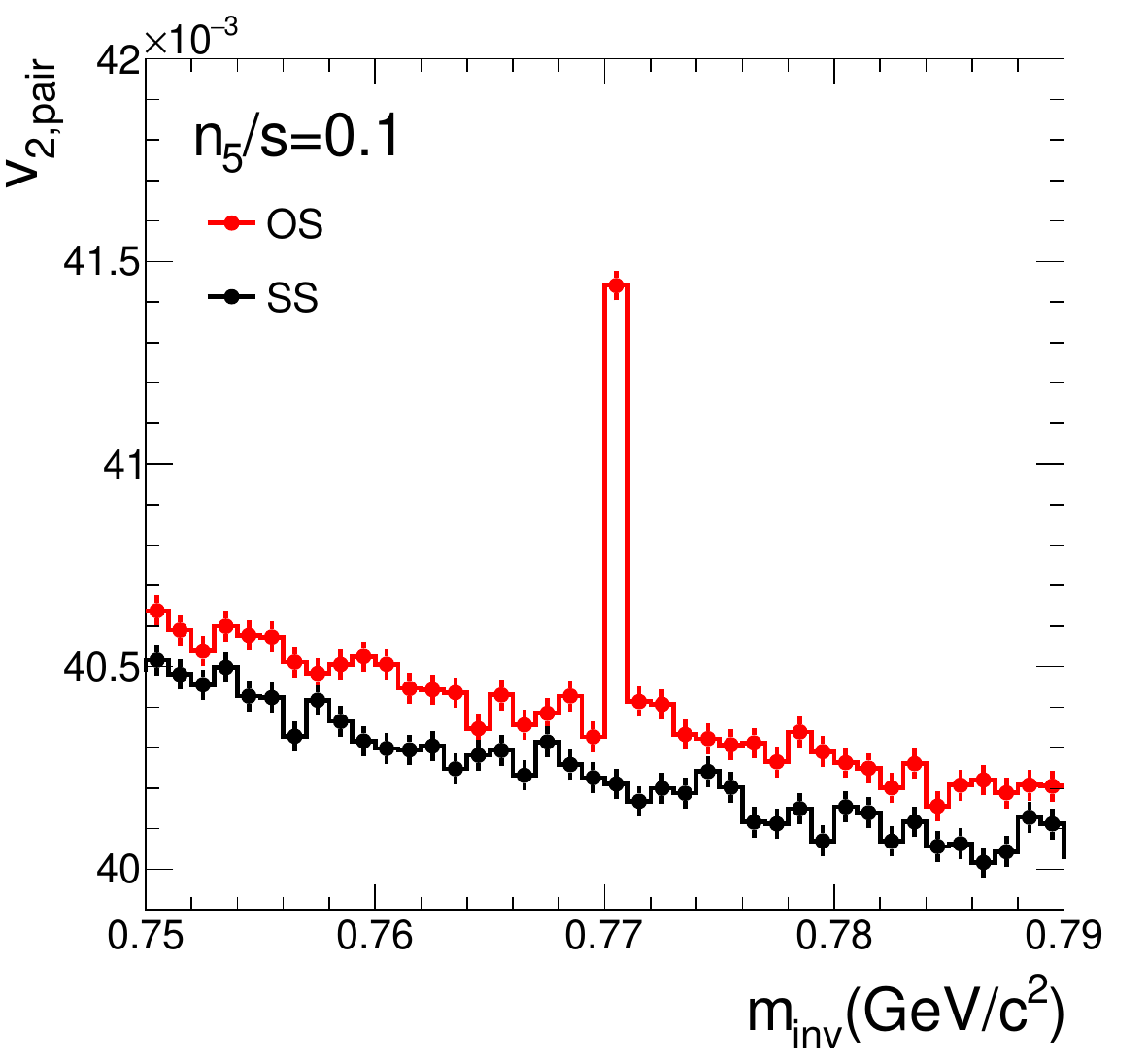}\hfill
\xincludegraphics[width=0.333\linewidth,pos=nwlow,label=\hspace{3.5cm}c)]{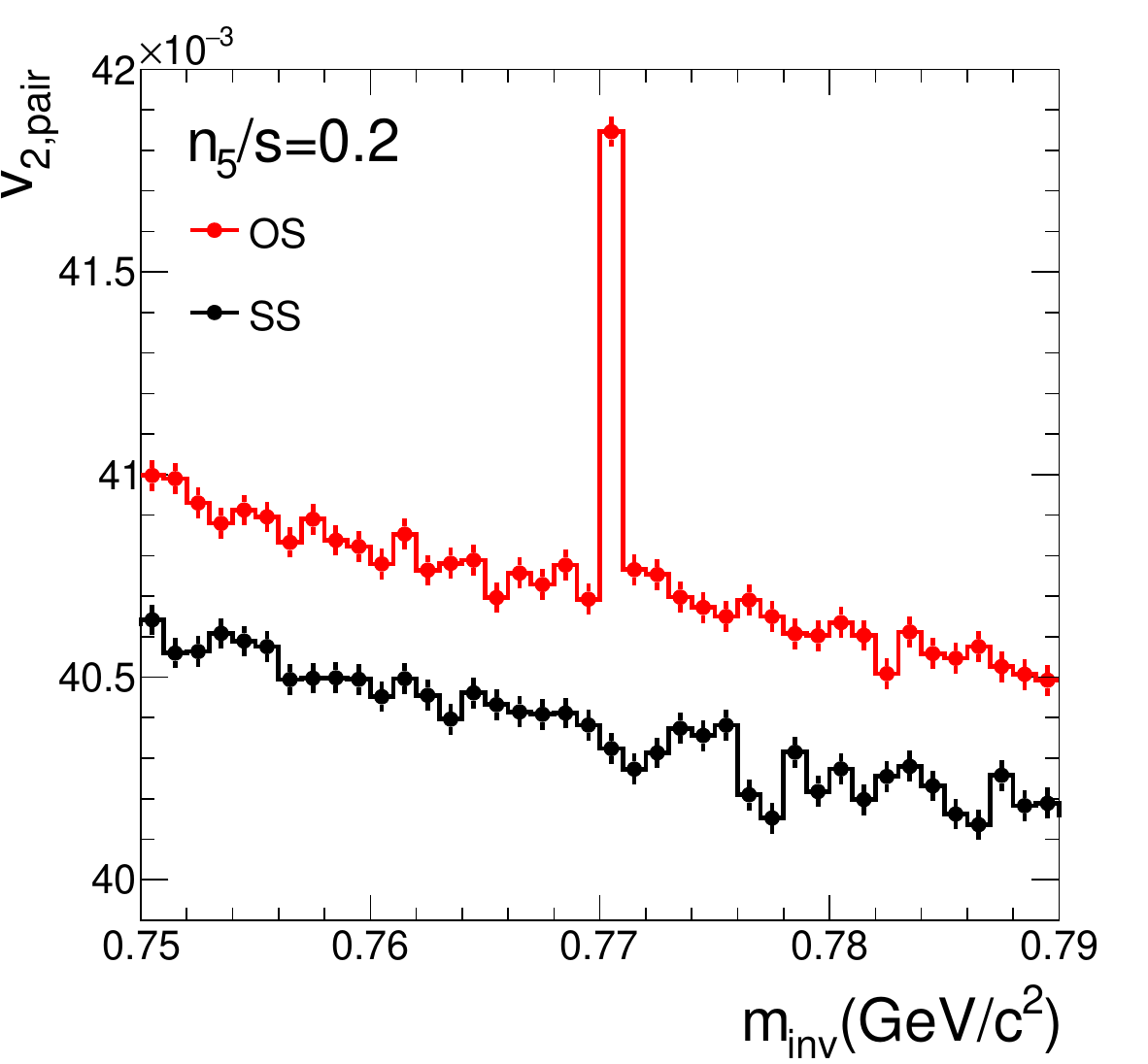}\hfill
\caption{Elliptic anisotropies $\vpair$ of OS and SS pairs with respect to the known impact parameter direction as functions of $\minv$, simulated by AVFD for $\ns=0$ (a), 0.1 (b), and 0.2 (c).}
\label{fig:v2res}
\end{figure*}

It is interesting to investigate the $\vres$ of those resonances, which can be obtained from elliptic anisotropies of OS pairs ($\vos$) and SS pairs ($\vsm$). The pair azimuthal anisotropies (i.e., $\vos$ and $\vsm$) are given by $\vpair \equiv \mean{\cos2\phi_\pair}$ where $\phi_\pair$  is the pair azimuthal angle relative to the RP. We focus on the $\rho$ resonance here because it is the dominant resonance contributing to OS pairs. The $\vos$ and $\vsm$ are shown in Fig.~\ref{fig:v2res} as functions of $\minv$ in the vicinity of the $\rho$ resonance peak for the three $n_{5}/s$ cases.

\begin{figure}[hbt]
\includegraphics[width=0.9\linewidth]{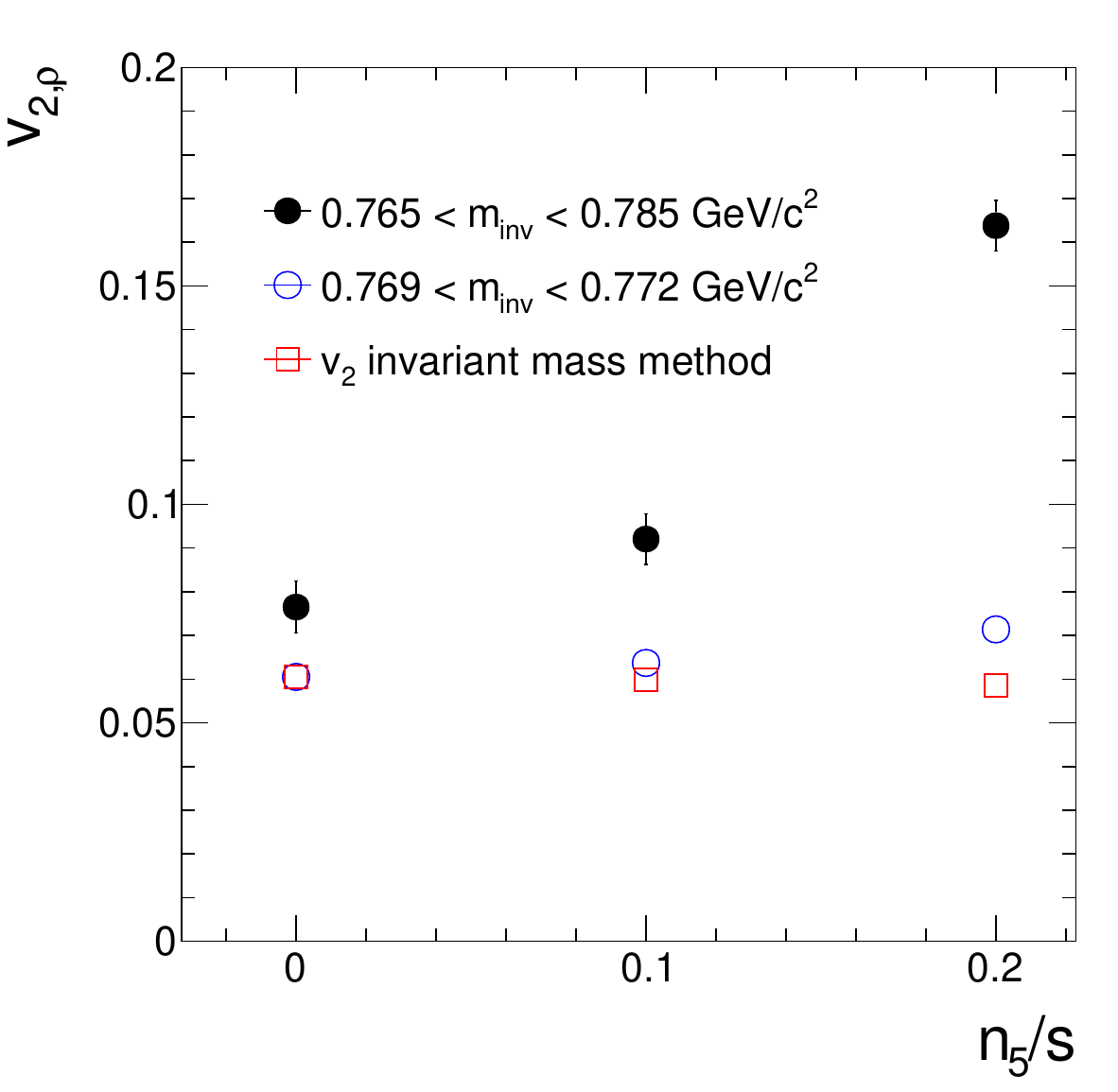}
\caption{The $\rho$ resonance $\vrho$ calculated by Eq.~(\ref{eq:v2res_ucla}) with a wide mass window $0.765<\minv<0.785$~\gevcc\ (black filled circles) and a narrow mass window $0.769<\minv<0.772$~\gevcc\ (blue open circles), plotted vs.~$\ns$. The $\vrho$ calculated by Eq.~(\ref{eq:v2res}) is shown in red squares.}
\label{fig:v2rho}
\end{figure}

In Ref.~\cite{Xu:2023elq} the resonance $\vres$ is calculated, after applying a $\minv$ window, by
\begin{equation}
    \vres = \frac
    {N_{\os} \mean{\cos2\phi_\os} - N_{\sm} \mean{\cos2\phi_\sm} }
    {N_{\os} - N_{\sm}}\,.
    \label{eq:v2res_ucla}
\end{equation}
This would work if $v_{2,\os}$ and $v_{2,\sm}$ are equal outside of the resonance invariant mass peak and $N_\sm$ represents the true number of background pairs underneath the resonance peak of $N_\os$ pairs. Neither of these two conditions is valid as seen from Figs.~\ref{fig:minv} and~\ref{fig:v2res}. Thus, the calculated $\vres$ by Eq.~\ref{eq:v2res_ucla} would be incorrect. This is particularly so if a wide $\minv$ window is applied to embrace the resonance peak, as shown in filled markers in Fig.~\ref{fig:v2rho} for the $\rho$ resonance as an example. The calculated $\vres$ increases with $\ns$ because of an increasing departure between $v_{2,\os}$ and $v_{2,\sm}$. With a narrow mass window, the issue becomes less significant  but non-vanishing as indicated by the open circles in  Fig.~\ref{fig:v2rho}.

Often resonance $\vres$ is analyzed by the so-called invariant mass method~\cite{Abelev:2008ae}, where the $\mean{\cos2\phi}$ is calculated as a function of $\minv$, and a fit is performed taking into account the $\minv$ dependence of the resonance signal to background ratio and assuming a background $v_2$ dependence on $\minv$ (typically a first- or second-order polynomial). In this case, only the OS information are needed if one assumes the pair multiplicity and $v_2$ background shapes to be interpolated from side bands. In the same spirit, we can use SS pairs to approximate those background shapes. We calculate the $\vres$ using the latter approach, namely by scaling the SS quantities up to match the OS's,
\begin{equation}
    \vres = \frac
    {N_{\os} \mean{\cos2\phi_\os} - r_N r_{v_2}N_{\sm} \mean{\cos2\phi_\sm} }
    {N_{\os} - r_N N_{\sm}}\,.
    \label{eq:v2res}
\end{equation}
Here, $r_N$ is the scaling factor to scale the SS multiplicity up to that of OS's, and $r_{v_2}$ is that to scale $v_{2,\sm}$ up to match  $v_{2,\os}$ in the non-resonance (sideband) regions.
The obtained $\vres$ is shown by the red squares in Fig.~\ref{fig:v2rho}. The $\vrho$ appears to be  constant over $\ns$, suggesting that influences from the CME on resonance elliptic anisotropies, if any, are small.

\section{Pair Elliptic Anisotropy Difference}
\begin{figure*}[hbt]
\xincludegraphics[height=6cm,pos=swHIHI,label=\hspace{1cm}a)]{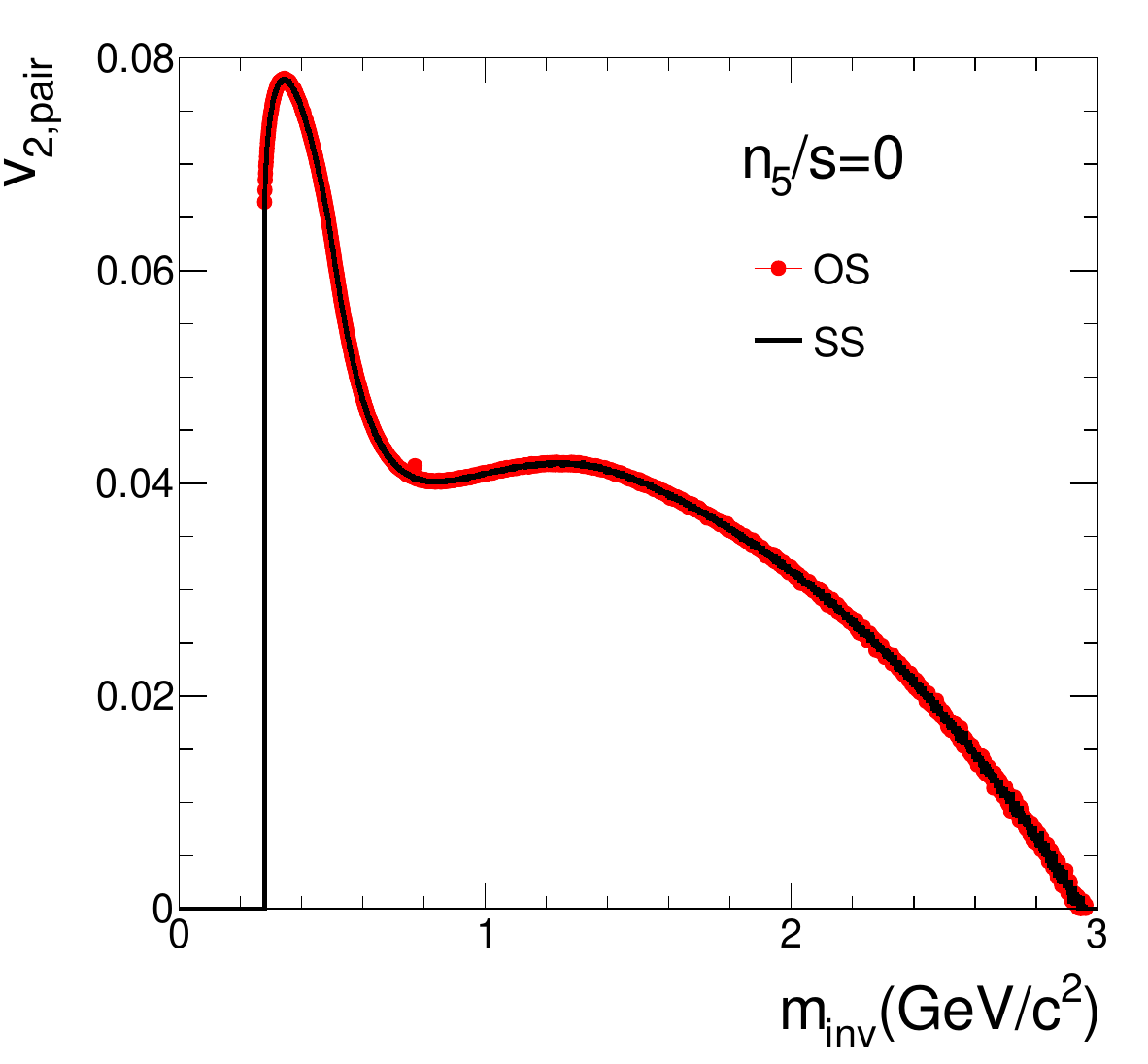}\hspace{1cm}
\xincludegraphics[height=6cm,pos=swHIGH,label=\hspace{1cm}b)]{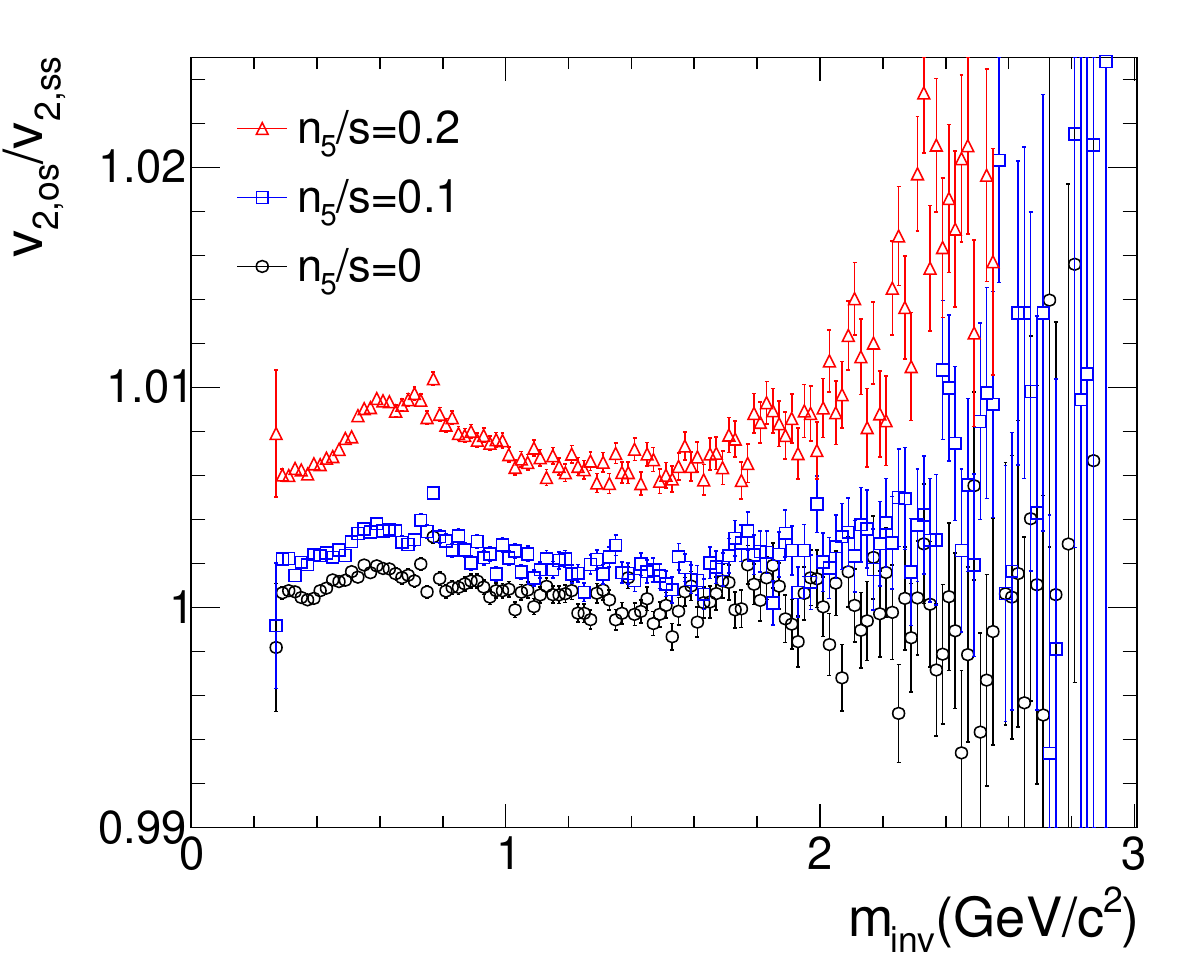}
\caption{(a) Elliptic anisotropies $\vpair$ of OS and SS pairs with respect to the known impact parameter direction as a function of $\minv$ over a wide range in $\minv$, simulated by AVFD for $\ns=0$.
(b) Ratio of OS and SS pair elliptic anisotropies, $\vos/\vsm$, as functions of $\minv$.}
\label{fig:v2minv}
\end{figure*}

\begin{figure}[hbt]
\includegraphics[width=0.9\linewidth]{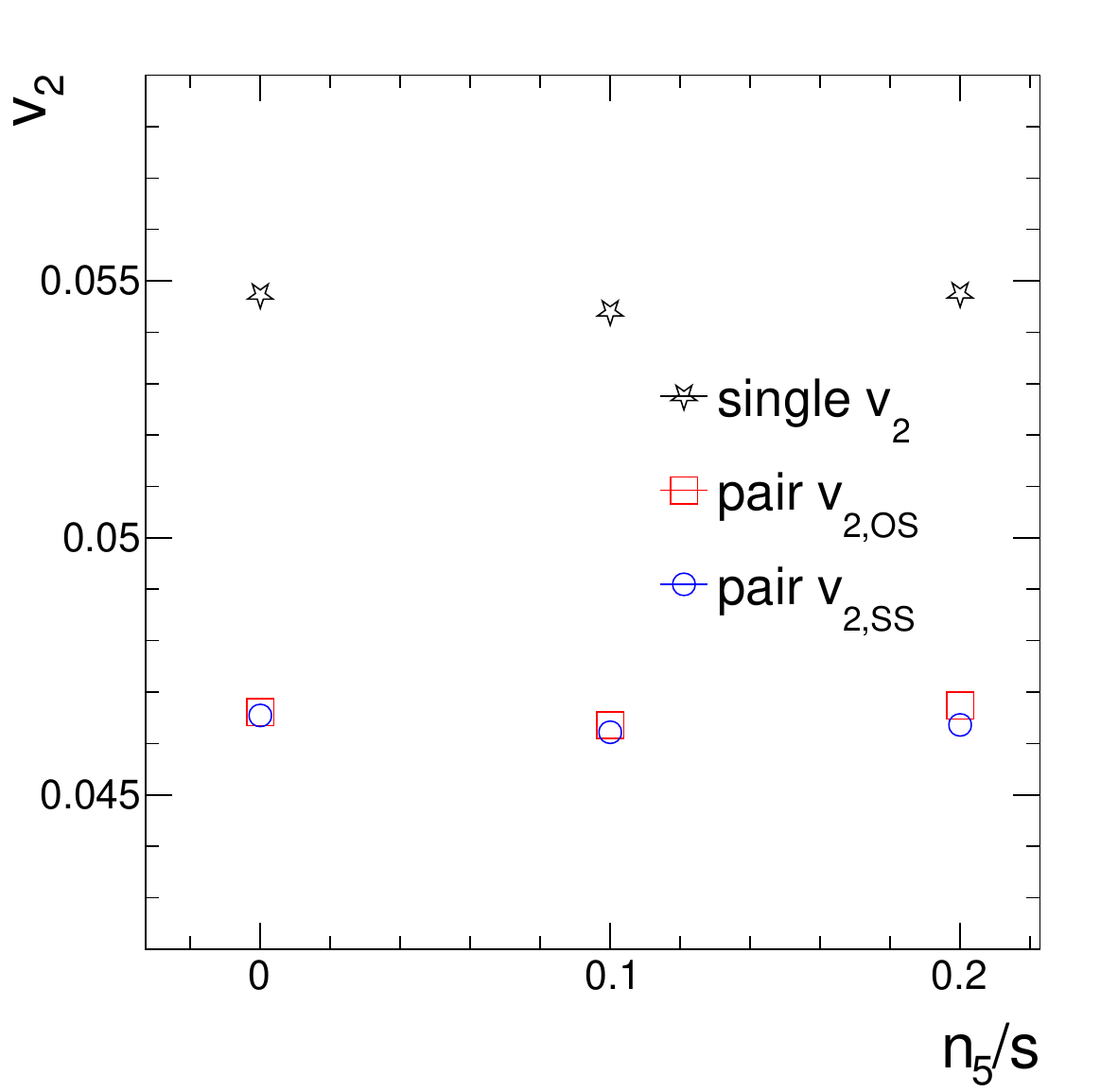}\hfill
\caption{Pair anisotropies of OS pairs $\vos$ (squares) and SS pairs $\vsm$ (circles) as functions of $n_{5}/s$, together with single particle $\vpi$ (stars). Statistical error bars are smaller than the marker size.}
\label{fig:v2pair}
\end{figure}

It is interesting to notice that the splitting between $\vos$ and $\vsm$ in Fig.~\ref{fig:v2rho} increases with increasing $\ns$. This is observed also in the entire region of $\minv$  as shown in Fig.~\ref{fig:v2minv}. In other words, the CME in AVFD leaves an imprint in this split. 
This has been pointed out by Ref.~\cite{Xu:2023elq}. It is therefore interesting to examine this split. To this end, we first obtain the  pair $\vos$ and $\vsm$ averaged over the entire $\minv$ range.
These are shown in Fig.~\ref{fig:v2pair}.
For comparison, the single particle $\vpi$ is also shown in Fig.~\ref{fig:v2pair}.

\begin{figure}[hbt]
\includegraphics[width=0.9\linewidth]{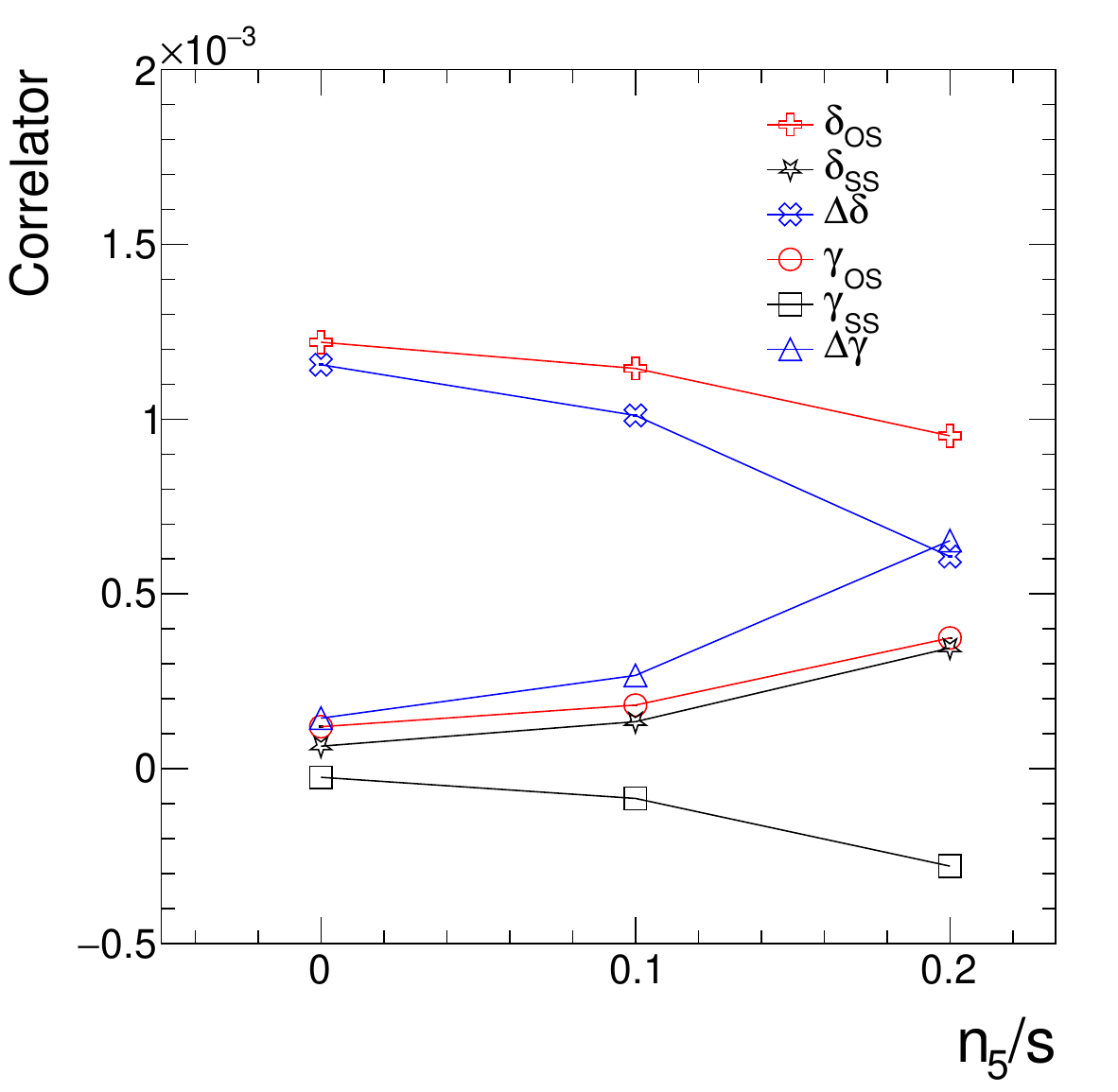}\hfill
\caption{Calculated $\gamma$ and  $\delta$ correlators for OS and SS pairs and the corresponding  $\Delta\gamma$ and $\Delta\delta$ correlators as functions of $n_{5}/s$ by AVFD. Statistical error bars are smaller than the marker size.}
\label{fig:gamma}
\end{figure}

To gain further insights, we make approximations to analytically examine the pair $\vpair$, following Ref.~\cite{Xu:2023elq}.
The pair elliptic anisotropy is by definition,
\begin{equation}
    \vpair=\mean{\frac{(p_{x,\alpha}+p_{x,\beta})^2-(p_{y,\alpha}+p_{y,\beta})^2}{(p_{x,\alpha}+p_{x,\beta})^2+(p_{y,\alpha}+p_{y,\beta})^2}}\,,
\end{equation}
where pairs can be distinguished between OS and SS pairs.
For equal $p_T$ pairs and under certain assumptions, it may be expressed into~\cite{Xu:2023elq}
\begin{eqnarray}
   \vpair &=& \mean{\frac{(\cos2\phi_\alpha+\cos2\phi_\beta)/2+\cos(\phi_\alpha+\phi_\beta)}{1+\cos(\phi_\alpha-\phi_\beta)}} \nonumber\\
   &\approx& \vpi(1-\delta) + \gamma\,,
   \label{eq:v2pair}
\end{eqnarray}
where $\delta \equiv \mean{\cos(\phi_\alpha-\phi_\beta)}$.
From this approxmination, the major component in $\vpair$ appears to be the single particle $\vpi$. Indeed, this is the case from AVFD simulation, as shown in Fig.~\ref{fig:v2pair}. The difference between $\vpair$ and $\vpi$ from the approximation is on the order of $\gamma$ and $\vpi\delta$.
For reference, we show in Fig.~\ref{fig:gamma} the $\gamma$ and $\delta$ quantities from the AVFD simulation. 
The trends in $\ns$ are opposite between OS and SS for both $\gamma$ and $\delta$ as expected from CME; they are not symmetric about zero because of non-CME background contributions. The trends in $\ns$ for each pair sign (OS or SS) between $\gamma$ and $\delta$ are opposite because CME contribution to $\delta\os$ (or $\delta\sm$) is of the opposite sign of that to $\gamma_\os$ (or $\gamma_\sm$).
The $\gamma$ and $\delta$ shown in Fig.~\ref{fig:gamma} are significantly smaller than the differences observed between $\vpair$ and $\vpi$ in Fig.~\ref{fig:v2pair}, so quantitatively,
Eq.~\ref{eq:v2pair} cannot describe  $\vpair$. This can be understood because gross assumptions are made in deriving Eq.~\ref{eq:v2pair}. For instance, for equal $p_T$ pairs, the $\vpair$ would  simply be $\mean{\cos(\alpha+\beta)}$ which is the $\gamma$ variable weighted by equal-$p_T$ pair probabilities.
Eq.~\ref{eq:v2pair} is helpful in illustrating that the pair $\vpair$  is on the same order of the single particle  $\vpi$. The pair $\vpair$ should not be confused with the two-particle anisotropy in  two-particle correlations which is approximately $\vpi^2$. However, Eq.~\ref{eq:v2pair} is not generally correct but only serves as a rough approximation. 

\begin{figure}[hbt]
\includegraphics[width=0.9\linewidth]{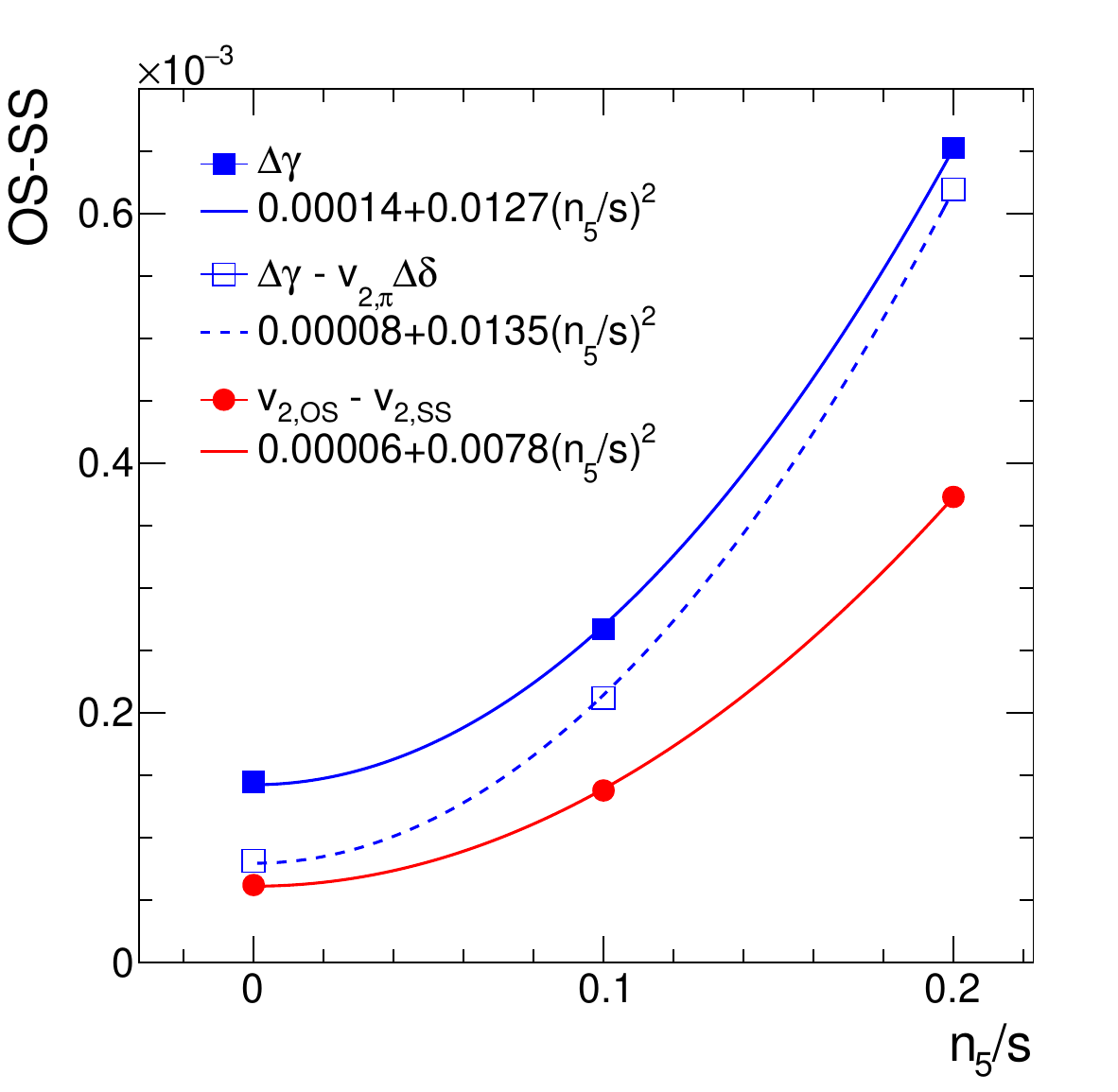}\hfill
\caption{Pair anisotropy difference $\Delta\vpair\equiv v_{2,\os}-v_{2,\sm}$ as functions of $n_{5}/s$, together with $\Delta\gamma$ and $\Delta\gamma-\vpi\Delta\delta$. The $n_{5}/s=0$ data are pure background, while at $n_{5}/s$=0.1 and 0.2 both CME signal and background are present.}
\label{fig:dv2}
\end{figure}

We are interested in the $v_2$ split between OS and SS pairs, 
\begin{equation}
    \Delta\vpair\equiv\vos-\vsm \,.
\end{equation}
Figure~\ref{fig:dv2} shows $\Delta\vpair$ as a function of $\ns$ in filled circles.
An increase in $\Delta\vpair$ with $\ns$ is observed. The red curve is a fit to $\Delta\vpair$, which is  dominated by the quadratic term $(\ns)^2$ and describes $\Delta\vpair$ well, suggesting that $\Delta\vpair$ is sensitive to CME. 
The nonzero $\Delta\vpair$ at $\ns=0$ suggests a background contribution to $\Delta\vpair$. 

From the approximation of Eq.~\ref{eq:v2pair}, one can obtain
%
%
\begin{widetext}
\begin{equation}
   \Delta\vpair \approx \Delta\gamma - \vpi\Delta\delta 
   \approx 2a_1^2(1+\vpi) + \frac{N_\res}{N_\os}
   \left[\mean{\cos(\phi_\alpha+\phi_\beta-2\phi_\res)}\vres -\mean{\cos(\phi_\alpha-\phi_\beta)}_\res\vpi \right]\,.  
   \label{eq:v2diff}
\end{equation}
\end{widetext}
It appears to be sensitive to both $\Delta\gamma$ and $\Delta\delta$, containing contributions from the CME and background.
For comparison, we superimpose $\Delta\gamma-\vpi\Delta\delta$ in Fig.~\ref{fig:dv2} in open squares.
Again the data points can not be qauntitatively described by Eq.~\ref{eq:v2diff} because of the  assumptions made in deriving the equations, but the agreement is better than for the individual $\vos$ and $\vsm$ in Fig.~\ref{fig:v2pair}. This is presumably because of cancellation of the inaccuracies in $\vos$ and $\vsm$ caused by the approximations.

The behavior of $\Delta\vpair$ is reminiscent of the $\Delta\gamma$ observable~\cite{Voloshin:2004vk} which is also composed of both CME signal and background and CME signal part is also quadratic in $\ns$. 
This is not surprising because the $\Delta\vpair$ observable is effectively as same as the $\Delta\gamma$ observable, with the arithmetic average angle $(\phi_\alpha+\phi_\beta)/2$ replaced by the ``$p_T$-weighted" one (i.e.~the pair azimuthal direction).
For comparison, we plot $\Delta\gamma$ in Fig.~\ref{fig:dv2} in filled squares, also with a quadratic fit. 

The three quantities, $\Delta\gamma$, $\Delta\vpair$, and $\Delta\gamma-\vpi\Delta\delta$, are all sensitive to the CME but are also all vulnerable to backgrounds. 
It appears that $\Delta\vpair$ and $\Delta\gamma-\vpi\Delta\delta$ may have a smaller background contribution than $\Delta\gamma$ has, presumably because of partial cancellations of backgrounds in $\Delta\gamma$ and $\Delta\delta$. 
One may even consider the quantity $\Delta\gamma-\kappa\vpi\Delta\delta$ with a $\kappa>1$ parameter~\cite{Adamczyk:2014mzf} to remove more background. 
However, the background contributions cannot be calculated a priori but have to be determined from data. Thus, all these observables are  similar and have qualitatively the same sensitivities to the CME signal and background.

\section{summary}
We have investigated, using the AVFD model, effects of the chiral magnetic effect (CME) on opposite-sign (OS) and same-sign (SS) pair-wise elliptic anisotropies, particularly on their difference $\Delta\vpair\equiv\vos-\vsm$. It is found that $\Delta\vpair$ increases quadratically with increasing axial current density $\ns$, as expected. It is also found that the $\Delta\vpair$ contains a background contribution. The CME signal and background in $\Delta\vpair$ are also difficult to separate, similar to those in the charge-dependent azimuthal correlator $\Delta\gamma$ observable. We conclude that the sensitivities of $\Delta\vpair$ and $\Delta\gamma$ (and $\Delta\gamma-\vpi\Delta\delta$) to the CME and background are similar, and the $\Delta\vpair$ observable probably does not bring in significantly additional insights beyond the $\Delta\gamma$ observable. 

We have also studied resonance elliptic anisotropies ($\vres$) as a function of $\ns$. While there may be potential contribution from the CME to $\vres$, it does not seem obvious and statistically significant to gain additional insights on the CME from resonance $\vres$.

\section*{Acknowledgments}
We thank Dr.~Jinfeng Liao, Dr.~Shuzhe Shi, and Dr.~Gang Wang for useful discussions.
This work was supported in part by the U.S.~Department of Energy (Grant No.~DE-SC0012910).

\bibliography{ref}
\end{document}